\documentclass{PoS}

\title{Status of hadronic light-by-light scattering and the muon $(g-2)$}

\ShortTitle{Status of HLBL for $(g-2)_{\mu}$ }

\author{\speaker{P. Masjuan}\\
        Grup de F\'{\i}sica Te\`orica, Departament de F\'{\i}sica, 
  Universitat Aut\`onoma de Barcelona, and Institut de F\'{\i}sica d'Altes Energies (IFAE), 
  The Barcelona Institute of Science and Technology (BIST), Campus UAB, E-08193 Bellaterra (Barcelona), Spain\\
        E-mail: \email{masjuan@ifae.es}}

\author{P. Sanchez-Puertas\\
        Faculty of Mathematics and Physics, Institute of Particle and Nuclear Physics, Charles University in Prague, V Hole\v{s}ovi\v{c}k\'ach 2, Praha 8, Czech Republic\\
        E-mail: \email{sanchezp@ipnp.troja.mff.cuni.cz}}

\abstract{In this talk we review the recent progress on the numerical determination of the Hadronic Light-by-Light contribution to the anomalous magnetic moment of the muon and we discuss the role of experimental data on the accuracy of its determination. Special emphasis on the main contribution, the pseudoscalar piece, is made. Gathering recent progress in the light-by-light scattering contribution we consider 
$a_{\mu}^{{\mathrm{HLBL}}} = (12.1\pm3.0)\times10^{-10}$ as a good summary of the state-of-the-art calculations which still claims for a $4\sigma$ deviation between theory and experiment for the $(g-2)_{\mu}$.}

\FullConference{The 15th International Conference on Flavor Physics \& CP Violation\\
		5-9 June 2017\\
		Prague, Czech Republic}

\begin{document}

\section{Introduction}
\label{intro}

The anomalous magnetic moment of the muon $(g-2)_{\mu}$ is one of the most accurately measured quantities in particle physics, and as such is a very promising signal of new physics if a deviation from its prediction in the Standard Model is found. 

The present experimental value for $a_{\mu}=(g-2)_{\mu}/2$, is given by $a_{\mu}^{\mathrm{EXP}}=116 592 09.1(6.3)\times10^{-10}$, as an average of $a_{\mu^+}=116 592 04(7.8)\times10^{-10}$ and $a_{\mu^-}=116 592 15(8.5)\times10^{-10}$ \cite{Bennett:2004pv,PDG2014}. Since statistical errors are the largest source of uncertainties, a new measurement with a precision of $1.6 \times 10^{-10}$ is being pursuit at FNAL \cite{Carey:2009zzb} and JPARC~\cite{JPARC}, using different experimental techniques.

At the level of the experimental accuracy, the QED contributions have been completed up to the fifth order ${\cal O}(\alpha_{em}^5)$, giving the QED contribution $11658471.885(4)\times10^{-10}$ \cite{Aoyama:2012wk}, when using the Rydberg constant and the ratio $m_{Rb}/m_e$ as inputs~\cite{PDG2014}. Also electroweak (EW) and hadronic contributions  in terms of the hadronic vacuum polarization (HVP) and the hadronic light-by-light scattering (HLBL) are necessary. The latter represents the main uncertainty in the Standard Model. The common estimates for QED, HVP, HLBL, and EW corrections are collected in Table~\ref{SMcont}. In this talk, we will update the HLBL contribution.

\begin{table}[htbp]
\begin{center}
\begin{tabular}{ccc}
Contribution  &  Result in $10^{-10}$ units & Ref.\\[5pt]
\hline
QED (leptons) & $11658471.885\pm 0.004$ & \cite{Aoyama:2012wk}\\
HVP (leading order) & $690.8\pm4.7$ & \cite{Hagiwara:2011af}\\
HVP$^{\mathrm{NLO+NNLO}}$ & $-8.7\pm0.1$& \cite{Hagiwara:2011af,Kurz:2014wya}\\
HLBL & $11.6 \pm4.0$ & \cite{Jegerlehner:2009ry}\\
EW & $15.4\pm0.1$ &\cite{Gnendiger:2013pva}\\
\hline
Total & $11659179.1\pm6.2$\\
\hline
\end{tabular}
\end{center}
\caption{Standard Model contributions to $(g-2)_{\mu}$.}
\label{SMcont}
\end{table}%

For the HLBL, two reference numbers can be found in the literature. The one quoted in Table~\ref{SMcont} $a_{\mu}^{{\mathrm{HLBL}}} = (11.6\pm4.0)\times10^{-10}$~\cite{Jegerlehner:2009ry} but also $(10.5\pm2.5)\times10^{-10}$~\cite{Prades:2009tw}. They both imply a discrepancy $\Delta a_{\mu} = a_{\mu}^{\mathrm{EXP}} - a_{\mu}^{\mathrm{SM}}=(30.0\pm8.8)\times10^{-10}$ of about  $3.5\sigma$. The overall HLBL contribution is twice the order of the present experimental error and a third of $\Delta a_{\mu}$. The striking situation then comes when the foreseen experiments (precision of $1.6\times 10^{-10}$) would imply the HLBL being a $6\sigma$ effect. On the light of such numbers we really need to understand the HLBL values and their errors since the goal is a HLBL within a $10\%$ uncertainty. 

The progress on the field is captured in at least three recent dedicated workshops on $(g-2)_{\mu}$~\cite{Czyz:2013zga,Benayoun:2014tra,Proceedings:2016bng} and a newly created $(g-2)_{\mu}$ theory initiative:\\
https://indico.fnal.gov/conferenceDisplay.py?confId=13795. 
In this letter, we focus our attention to the HLBL. For prospects in reducing the error for the HVP we refer to the talks by A. Keshavarzi and B.Malaescu in the aforementioned g-2 initiative webpage as well as the contribution by V. Druzhinin in this proceedings.

The results here described update those reported in Ref.~\cite{Masjuan:2014rea}.

\begin{table*}[htbp]
\caption{The HLBL and its different contributions from different references and methods, representing the progress on the field and the variety of approaches considered. $\dag$ indicates used from a previous calculation. Units of $10^{-11}$.}
\begin{center}
{\small
\begin{tabular}{llccccl}
Group& HLBL & $\pi,K$ loop & PS & Higher spin & Quark loop & Method\\
\hline
BPP~\cite{Bijnens:1995cc} &$+83(32)$ & $-19(13)$ &  $+85(13)$ &  $-4(3)$ &  $+21(3)$ & \textrm{ENJL, '95\, '96\, '02}\\  
HKS~\cite{Hayakawa:1995ps} &$+90(15)$ & $-5(8)$ &  $+83(6)$ &  $+1.7(1.7)$ &  $+10(11)$& \textrm{LHS, '95\, '96\, '02}  \\  
KN~\cite{Knecht:2001qf} &$+80(40)$ & & $+83(12)$ &  & &\textrm{Large $N_c$+$\chi$PT, '02}\\
MV~\cite{Melnikov:2003xd} &$+136(25)$ & $0(10)$ &  $+114(10)$ &  $+22(5)$ &  $0$&\textrm{Large $N_c$+$\chi$PT, '04} \\    
JN~\cite{Jegerlehner:2009ry} &$+116(40)$ & $-19(13)\dag$ &  $+99(16)$ &  $+15(7)$ &  $+21(3)\dag$&\textrm{Large $N_c$+$\chi$PT, '09}\\  
PdRV~\cite{Prades:2009tw} &$+105(26)$ & $-19(19)$ &  $+114(13)$ &  $+8(12)$ &  $0$&\textrm{Average, '09}\\  
HK~\cite{Hong:2009zw} &$+107$ &  & $+107$ &  && \textrm{Hologr. QCD, '09} \\  
DRZ~\cite{Dorokhov:2014iva} &$+168(13)$ &  &$+59(9)$ &  &$+110(9)$& \textrm{Non-local q.m., '11} \\  
EMS~\cite{Masjuan:2012wy,Masjuan:2012qn,Escribano:2013kba} &$$ &  $$ &  $+90(7)$ &   $$ &  $$ &\textrm{Pad\'e-data'13}\\ 
EMS~\cite{Masjuan:2012sk,Escribano:2013kba} &$$ &  $$ &  $+88(4)$ &   $$ &  $$ &\textrm{Large $N_c$ , '13} \\   
GLCR~\cite{Roig:2014uja} &$$ & $$ &  $+105(5)$ &   $$ &  $$& \textrm{Large $N_c$+$\chi$PT, '14} \\  
J~\cite{Jegerlehner:2015stw} &$$ & $$ &  $$ &   $+8(3)_{\rm axial}$ &  $$& \textrm{Large $N_c$+$\chi$PT, '15} \\  
BR~\cite{Bijnens:2016hgx}&$$ & $-20(5)_{\pi {\rm \, only} }$ &  $$ &   $$ &  $$& \textrm{Large $N_c$+$\chi$PT, '16} \\  
MS~\cite{Masjuan:2017tvw} &$$ & $$ &  $+94(5)$ &   $$ &  $$&\textrm{Pad\'e-data '17}\\ 
CHPS~\cite{Colangelo:2017qdm} &$$ & $-24(1)_{\pi {\rm \, only} }$ &  $$ &   $$ &  $$& \textrm{Disp Rel, '17} \\  
\end{tabular}
}
\end{center}
\label{T2}
\end{table*}%

\section{Dissection of the HLBL and potential issues}

The HLBL cannot be directly related to any measurable cross section and requires knowledge of QCD at all energy scales. Since this is not known yet, one needs to rely on hadronic models to compute it. Such models introduce systematic errors which are difficult to quantify. Using the large-$N_c$ and the chiral counting, de Rafael proposed~\cite{deRafael:1993za} to split the HLBL into a set of different contributions: pseudoscalar exchange (PS, dominant~\cite{Jegerlehner:2009ry,Prades:2009tw}), charged pion and kaon loops, quark loop, and higher-spin exchanges (see Table~\ref{T2}, notice the units of $10^{-11}$). The large-$N_c$ approach however has at least two shortcomings: firstly, it is difficult to use experimental data in a large-$N_c$ world. Secondly, calculations carried out in the large-$N_c$ limit demand an infinite set of resonances. As such sum is not known, one truncates the spectral function in a resonance saturation scheme, the Minimal Hadronic Approximation (MHA)~\cite{Peris:1998nj}. The resonance masses used in each calculation are then taken as the physical ones from PDG~\cite{PDG2014} instead of the corresponding masses in the large-$N_c$ limit. Both problems might lead to large systematic errors not included so far~\cite{Masjuan:2012wy,Masjuan:2007ay,Masjuan:2012sk,Escribano:2013kba,Masjuan:2017tvw}, large and difficult to estimate. Results obtained under such assumptions are quoted as Large $N_c$+$\chi$PT in the last column of Table~\ref{T2}.

Actually, most of the results in the literature follow de Rafael's proposal (see Refs.~\cite{Hayakawa:1997rq,Bijnens:1995cc,Knecht:2001qg,Blokland:2001pb,Melnikov:2003xd,Dorokhov:2008pw,Nyffeler:2009tw,Jegerlehner:2009ry,Prades:2009tw,Hong:2009zw,Cappiello:2010uy,Kampf:2011ty,Masjuan:2012wy,Bijnens:2016hgx,Escribano:2013kba,Roig:2014uja,Dorokhov:2015psa,Colangelo:2017qdm}, including full and partial contributions to $a_{\mu}^{\textrm{HLBL}}$) finding values for $a_{\mu}^{\textrm{HLBL}}$ between basically $6 \times 10^{-10}$ and up to almost $14 \times 10^{-10}$.  

Such range almost reaches ballpark estimates based on the Laporta and Remiddi (LR)~\cite{Laporta:1992pa} analytical result for the heavy quark contribution to the LBL. The idea in such ballparks is to extend the perturbative result to hadronic scales low enough for accounting at once for the whole HLBL. The free parameter is the quark mass $m_q$. The recent estimates using such methodology~\cite{Pivovarov:2001mw,Erler:2006vu,Boughezal:2011vw,Greynat:2012ww,Masjuan:2012qn} found $m_q \sim 0.150 - 0.250$ GeV after comparing the particular model with the HVP. The value for the HLBL is higher than those shown in Table~\ref{T2}, around $a_{\mu}^{\mathrm{HLBL}}=12 - 17\times 10^{-10}$, which seems to indicate that the subleading pieces of the standard calculations seems to be non-negligible.

As we said, the Jegerlehner and Nyffeler review~\cite{Jegerlehner:2009ry} together with the \emph{Glasgow consensus} written by Prades, de Rafael, and Vainshtein~\cite{Prades:2009tw} represent, in our opinion, the two reference numbers. They agree well since they only differ by few subtleties. For the main contribution, the pseudoscalar, one needs a model for the pseudoscalar Transition Form Factor (TFF). They both used the model from Knecht and Nyffeler~\cite{Knecht:2001qf} based on MHA, but differ on how to implement the high-energy QCD constrains coming from the VVA Green's function. In practice, this translates into whether the piece contains a pion pole or a pion exchange; moreover, the first allegedly requires excluding the quark loop, whereas the latter defend its inclusion. In practice, the former would imply that the exchange of heavier pseudoscalar resonances (6th column in Table~\ref{T2}) is effectively included in PS~\cite{Melnikov:2003xd}, while the latter demands to take them into account separately. The other difference is whether the errors are summed linearly~\cite{Jegerlehner:2009ry} or in quadrature~\cite{Prades:2009tw}. All in all, even though the QCD features for the HLBL are well understood~\cite{Jegerlehner:2009ry,Prades:2009tw}, the details of the particular calculations are important to get the numerical result to the final required precision. Considering the drawback drawn here, we think we need more calculations, closer to experimental data if possible.

Dispersive approaches \cite{Colangelo:2015ama,Colangelo:2017qdm} rely on the splitting of the former LBL tensor into several pieces according to low-energy QCD, which most relevant intermediates states are selected according to their masses~\cite{deRafael:1993za,Kinoshita:1984it}; 
see Ref.~\cite{Colangelo:2017qdm} for recent advances. Up to now, only a subleading piece has been computed, cf. Table~\ref{T2}. An advantage we see in this approach is that by decomposing the LBL tensor in partial waves, a single contribution may incorporate pieces that were separated so far, avoiding potential double counting. The example is the $\gamma \gamma \to \pi \pi$ which includes the two-pion channel, the pion loop, and scalar and tensor contributions. A complete and model-independent treatment would require coupled channel formalism, not developed so far, and a matching to the high-energy region yet to be included. So by now, the calculations are not yet complete, and not yet ready to be added to the rest of contributions.

Finally, for the first time, there have been different proposals to perform a first principles evaluation by using lattice QCD~\cite{Blum:2014oka}. They studied a non-perturbative treatment of QED which later on was checked against the perturbative simulation. With that spirit, they considered that a QCD+QED simulation could deal with the non-perturbative effects of QCD for the HLBL. Whereas yet incomplete and with some progress still required, promising advances have been reported already~\cite{Blum:2014oka}.

\section{The role of the new experimental data on the HLBL}

The main obstacle when using experimental data is the lack of them, specially on the doubly virtual TFF~\cite{Sanchez-Puertas:2017sih}. Fortunately, data on the TFF when one of the photons is real is available from different collaborations, not only for $\pi^0$ but also for $\eta$ and $\eta'$. It is common to factorize the TFF, i.e., $F_{P\gamma^*\gamma^*}(Q_1^2,Q_2^2) = F_{P\gamma^*\gamma}(Q_1^2,0) \times F_{P\gamma \gamma^*}(0,Q_2^2)$, and describe it based on a rational function. A further refinement includes a modification of its numerator due to the high-energy QCD constraints~\cite{Knecht:2001qf}. Although the high-energy region of the model is not very important, it still contributes around  $20\%$. More important is the double virtuality, especially if one uses the same TFF model (as it should) for predicting the $\pi^0 \to e^+e^-$ decay. Current models cannot accommodate its experimental value (see~\cite{Masjuan:2015lca}) which call for a new--- more precise--- measurement. The worrisome fact is that modifying the model parameters to match such decay and going back to the HLBL, would result in a dramatic decrease of the HLBL value~\cite{Masjuan:2015lca}.

While the HLBL requires knowledge at all energies, it is condensed in the $Q^2$ region from $0$ to $2$ GeV$^2$, in particular above around $0.5$ GeV$^2$. Therefore a good description of TFF in such region is very important. Such data are not yet available, but any model should reproduce the available one. Therefore, any model relies on extrapolation from the medium- and high-energy region---where data is available---to the low-energy one, which is clearly a model-dependent procedure.  That is why the authors of \cite{Masjuan:2012wy,Masjuan:2012qn,Escribano:2013kba,Masjuan:2017tvw}, in contrast to other approaches, did not used data directly but the low-energy parameters (LEP) of the Taylor expansion for the TFF and reconstructed it \textit{via} the use of Pad\'e approximants. As demonstrated in Ref.~\cite{Masjuan:2017tvw}, the pseudoscalar TFF driving the PS contributions to the HLBL is Stieltjes functions (or more precisely, a rational function of Stieltjes type) for which the convergence of the Pad\'e approximants sequence is guaranteed in advanced. As such, a comparison between two consecutive elements in this sequence estimates the systematic error yield by the method. In other words, Pad\'e approximants for the TFFs take full advantage of analyticity and unitary of these functions to correctly extrapolate low- and high-energy regions. 

The LEPs certainly know about all the data at all energies and as such incorporates all our experimental knowledge at once. This procedure implies a model-independent result together with a well-defined way to ascribe a systematic error. It is a procedure that can be considered an \emph{approximation}, not an \emph{assumption}  (as such, our procedure does apply beyond the large-$N_C$ limit of QCD). The LEPs were obtained in~\cite{Masjuan:2012wy} for the $\pi^0$, in~\cite{Escribano:2015nra} for the $\eta$-TFF and in \cite{Escribano:2015yup} for the $\eta'$-TFF, taking into account the $\eta-\eta'$ mixing \cite{Escribano:2015yup,Bickert:2016fgy} (in addition, the relevant $\eta-\eta'$ mixing parameters were determined there and represent \emph{the} state-of-the-art) and the determinations of the double virtual $\pi^0$~\cite{Masjuan:2015lca} and $\eta,\eta'$~\cite{Masjuan:2015cjl} TFFs. Ref.~\cite{Masjuan:2017tvw} collects the most updated results for the space- and time-like TFF together with $\gamma \gamma$ decays from 13 different collaborations, and yields the most updated and precise pseudoscalar contribution to the HLBL. The HLBL value from such approach is quoted in Table~\ref{T2} under EMS and under MS after the double virtual $\pi^0, \eta, \eta'$-TFF were extracted from pseudoscalar decays into a lepton pair~\cite{Masjuan:2015lca,Masjuan:2015cjl}.

The new pseudoscalar-pole contribution obtained with the Pad\'e method yields $a_{\mu}^{{\mathrm{HLBL,PS}}} = (9.4\pm0.5)\times10^{-10}$~\cite{Masjuan:2017tvw}, which agrees very well with the \emph{old reference} numbers but with an error reduced by a factor of 3. Adding to this quantity the $\pi$ loop $(2.0\pm0.5)\times10^{-10}$ from~\cite{Bijnens:2016hgx}, the axial contribution $(+0.8\pm0.3)\times10^{-10}$ from~\cite{Jegerlehner:2015stw}, the scalar contribution $(-0.7\pm0.7)\times10^{-10}$ and the quark loop from~\cite{Bijnens:1995cc}, and the NLO estimate $(+0.3\pm0.2)\times10^{-10}$ from~\cite{Colangelo:2014qya}, the HLBL reads:
\begin{eqnarray}\label{result}
 a_{\mu}^{{\mathrm{HLBL}}} &= (9.9\pm1.1)\times10^{-10} \quad \quad {\rm (errors \,\, in \,\,quadrature)}\\
 a_{\mu}^{{\mathrm{HLBL}}} &= (9.9\pm2.5)\times10^{-10} \quad \quad {\rm (errors \,\,linearly\,\, added)}\, .
 \end{eqnarray}

The discussion not yet settled is whether one should consider a pseudoscalar-pole or a pseudoscalar-exchange contribution, which is the main difference between the two \emph{old reference} results, see the recent discussion in Ref.~\cite{Masjuan:2017tvw}. If instead of a pseudsocalar pole, one would consider the pseudoscalar exchange, Ref.~\cite{Masjuan:2017tvw} tells us that the result MS from Table~\ref{T2} would grow up to $a_{\mu}^{{\mathrm{HLBL,PS}}} = (13.5\pm1.1)\times10^{-10}$. Then, summing up the rest of the contributions (without the quark loop which is effectively included in the PS exchange), the HLBL reads:
\begin{eqnarray}\label{result2}
 a_{\mu}^{{\mathrm{HLBL}}} &= (12.1\pm1.5)\times10^{-10} \quad \quad {\rm (errors \,\, in \,\,quadrature)}\\
 a_{\mu}^{{\mathrm{HLBL}}} &= (12.1\pm3.0)\times10^{-10} \quad \quad {\rm (errors \,\,linearly\,\, added)}\, .
 \end{eqnarray}

While still marginally compatible, the results from Eqs.~(\ref{result}) and~(\ref{result2}) indicate once more that the role of the dismissed pieces in the standard pseudoscalar-pole framework seems to be as important as subleading contributions which are much larger than the desired global $10\%$ precision. This fact requires, of course, further calculations and focus on missing pieces now that the dominant one is well under control.  

In conclusion, the new experimental data and the correct high-energy constraints used to update the pseudoscalar contribution in Ref.~\cite{Masjuan:2017tvw} seem to reveal larger contributions from pseudsocalar mesons, meaning that the modeling of the TFF is more important than expected. Also, systematic errors due to both chiral and large-$N_c$ limits are important and difficult to evaluate, but PAs can help. Lattice QCD seems promising but only in the long run. Dispersion relations are useful at low energies for including subleading terms and potentially avoiding double-counting in particular diagrams. However,  a consensus will be needed in order to combine such results with those from other contributions, and the matching to high-energies (very important pursuing precision as we already discussed) remains still unsettled. On top of this, the ballpark predictions coincide on drawing scenarios with larger values, indicating in our opinion the need to better understand the process from a global point of view.

 \section*{Acknowledgements}
P.M is supported by CICYTFEDER-FPA2014-55613-P, 2014-SGR-1450, the CERCA Program/Generalitat de Catalunya, and the Secretaria d'Universitats i Recerca del Departament d'Empresa i Coneixement de la Generalitat de Catalunya. P.S.P. is supported by the Czech Science Foundation (grant no. GACR 15-18080S).

\end{document}